\documentclass[%
reprint,
amsmath,amssymb,
prl
]{revtex4-1}

\usepackage{mathtools} %extension package to amsmath, load amsmath implicitly
\usepackage{bm}% bold math

\usepackage[usenames,dvipsnames]{xcolor}
\usepackage{ulem} %use underline with \uline{...}
\usepackage{comment}

\usepackage{graphicx}
\usepackage{epsfig}
\usepackage{subfigure}

\usepackage{hyperref}% add hypertext capabilities

\begin{document}
	
	\title{Unconventional Pressure Dependent Interorbital and Interlayer Doping in Superconducting Nickelates}
	
	\author{Y. N. Huang}
	\affiliation{Department of Physics, Zhejiang University of Science and Technology, Hangzhou 310023, People’s Republic of China}
	
	\author{David J. Singh}
	\email{david.joseph.singh@gmail.com}
	\affiliation{Department of Physics and Astronomy, University of Missouri, Columbia, MO 65211-7010, USA}
	
	\date{\today}

	\begin{abstract}
		The discovery of nickelate superconductivity provided the first example of a non-copper-based material with superconductivity strongly analogous to the cuprates, but recent findings raise questions and inconsistencies around the electron counts and doping phase diagrams. We show using superconducting La$_4$Ni$_3$O$_{10}$
that there are unconventional interlayer and interorbital intrinsic doping effects that render the $d_{x^2-y^2}$ orbital occupation similar to the cuprates. The results enable a consistent framework for nickelate superconductivity, while maintaining the connection between cuprate and nickelate superconductors.  
	\end{abstract}
	
	\maketitle

	Understanding high temperature (high-$T_c$) cuprate superconductivity\cite{10.1007/BF01303701} is one of the most important problems in condensed matter physics. Although the mechanism of superconductivity has remained elusive, research since 1986 has revealed many features of the cuprates and their superconductivity. Much of the understanding is related to phase diagrams, particularly the complex behavior of the electronic structure, transport, and superconductivity, as well as the emergence of competing phases as a function of doping, i.e., the electron counts in the Cu $d$-orbitals, especially the deviation from half-filling of the $d_{x^2-y^2}$ orbital. The recent discoveries of nickelate superconductivity provided the first example of a Cu-free material showing superconductivity with the key characteristics of the cuprates\cite{10.1038/s41586-019-1496-5, 10.1038/s41467-020-19908-1, 10.1038/s41563-023-01766-z, 10.1126/sciadv.abl9927}. The commonalities and differences from the cuprates hold the promise of providing crucial insights into the seemingly common superconductivity of these materials.
	
	The initial discovery of nickelate superconductivity was in the doped infinite layer
$R$NiO$_2$ ($R$=rare earth) system\cite{10.1038/s41586-019-1496-5}.
The nominal valence of Ni in the stoichiometric material is Ni$^+$,
with the same $d$-electron count as the Cu$^{2+}$ in the high-$T_c$ parent La$_2$CuO$_4$.
Similar to the cuprates, NdNiO$_2$ is not superconducting but is an insulator,
while superconductivity emerges with doping by Sr.
This would correspond to hole doping similar to the (La,Sr)$_2$CuO$_4$ system.
This was followed by the discovery of superconductivity with $T_c = 80$ K under pressure in the bilayer compound
La$_3$Ni$_2$O$_7$ with nominal valence Ni$^{+2.5}$ and without chemical dopants\cite{10.1038/s41586-023-06408-7}.
This was understood in terms of accommodation of some holes in the Ni $d_{z^2}$ orbital
in addition to the $d_{x^2-y^2}$ orbital\cite{10.1038/s41586-023-06408-7,10.1038/s41467-024-48701-7}.
This involvement of two distinct $d$ orbitals on the active Ni site is distinct from the self-doped cuprates
such as YBa$_2$Cu$_3$O$_7$ \cite{10.1103/PhysRevLett.58.908} and
Bi$_2$Sr$_2$CaCu$_2$O$_8$ \cite{10.1126/science.239.4843.1015},
where the CuO$_2$ planes are doped by other layers,
i.e., the chain-Cu layer in YBa$_2$Cu$_3$O$_7$ \cite{10.1126/science.255.5040.46}
and Bi-O and Tl-O bands in the Bi- and Tl-based cuprates\cite{10.1103/PhysRevLett.60.1665,10.1016/0921-4534(92)90526-I,10.1038/nature01981}. The consequence is that more complex superconducting states other than the $d$-wave $x^2-y^2$ order of the cuprates become possible\cite{10.1103/PhysRevB.108.L201108}. 
	Finally, superconductivity has also been reported in the three-layer compound La$_4$Ni$_3$O$_{10}$ under pressure without chemical doping\cite{10.1007/s11433-023-2329-x,10.1088/0256-307X/41/1/017401,10.48550/arXiv.2311.07423,10.1038/s41586-024-07553-3,10.48550/arXiv.2405.19880}, 
	although no superconductivity was found in polycrystalline samples\cite{10.1016/j.jmst.2023.11.011}.
	Nonetheless, experiments show strong dependence of the superconducting
behavior on the precise O stoichiometry indicating the importance of electron count\cite{10.48550/arXiv.2405.19880}.
	From a structural point of view, this compound can be considered
intermediate between the bilayer La$_3$Ni$_2$O$_7$
and the infinite layer LaNiO$_2$.
The nominal Ni valence in La$_4$Ni$_3$O$_{10}$ is also different, Ni$^{+2.67}$.
The compound also differs from the other two in that it has two different Ni sheets. 
	This provides an additional degree of freedom that may help unravel the puzzle of doping of the superconductors.
	These experimental findings have motivated several theoretical proposals for the superconductivity \cite{10.48550/arXiv.2402.05085,10.48550/arXiv.2404.16600,10.48550/arXiv.2405.04340,10.48550/arXiv.2404.09162,10.48550/arXiv.2402.05285,10.48550/arXiv.2402.07902,10.48550/arXiv.2402.07196,10.48550/arXiv.2405.00092,10.48550/arXiv.2402.05447,10.1088/1361-648X/ad512c,10.1103/PhysRevB.109.235123,10.1103/PhysRevB.109.144511}.

Here we use an analysis of pressure dependent first principles calculations and find an unconventional pressure dependent doping of this material. This includes an interlayer component with transfer of carriers
between the outer and inner layer Ni and an interorbital component involving the interplay
of the $d_{x^2-y^2}$ and $d_{z^2}$ orbitals.
	
\begin{figure}[htbp]
		\centering
		\includegraphics[width=0.6\columnwidth]{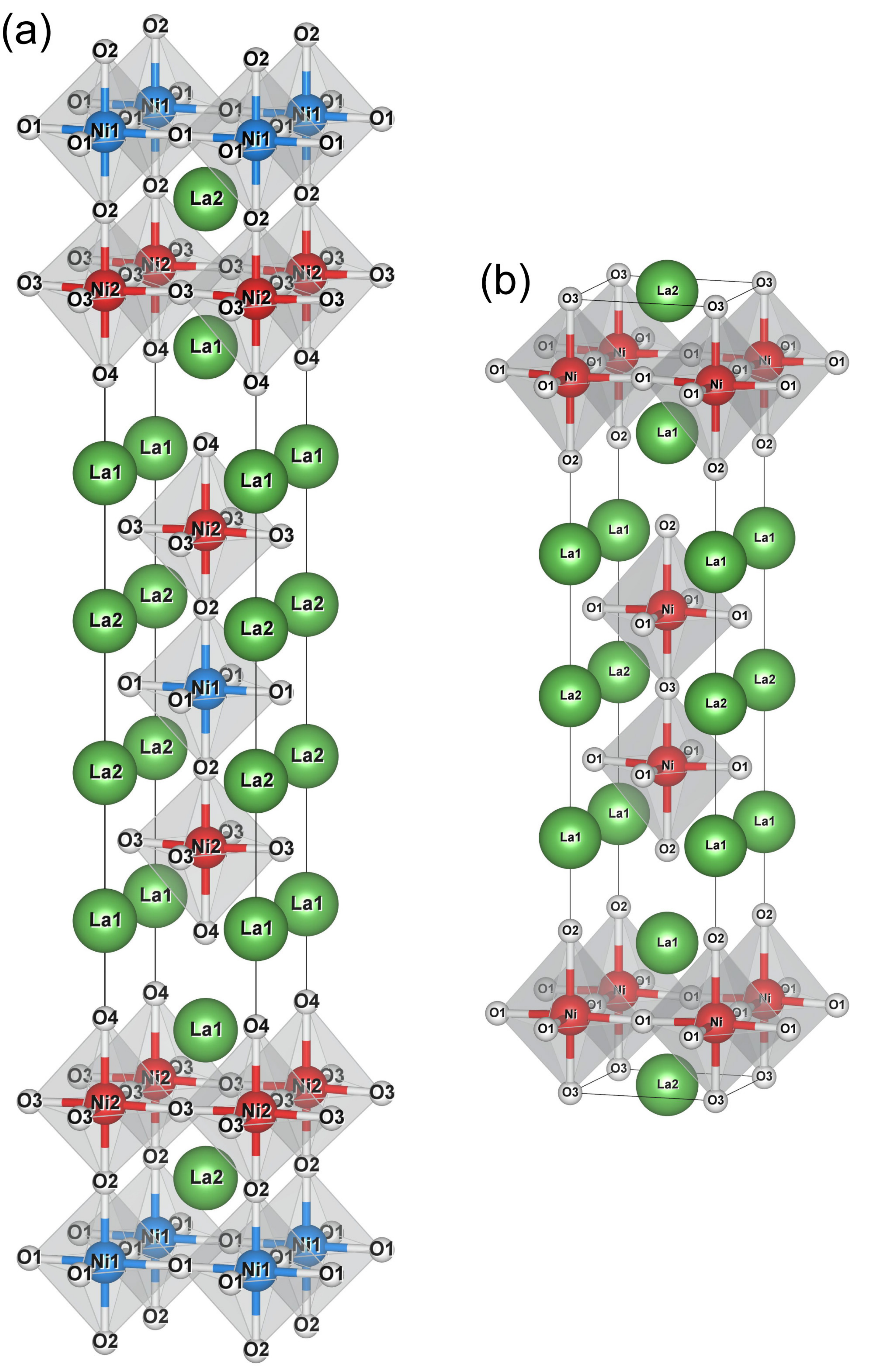}
		\caption{
			Crystal structure of (a) La$_4$Ni$_3$O$_{10}$ and (b) La$_3$Ni$_2$O$_7$ in $I4/mmm$ symmetry. We label nonequivalent atoms with respect to I4/mmm symmetry
		}
		\label{fig:4310_327_structure}
	\end{figure}

	La$_3$Ni$_2$O$_7$\cite{10.1021/jacs.3c13094} maintains $Amam$ symmetry under ambient conditions.
Upon cooling, this compound transitions to $Fmmm$ symmetry and, below 40 K,
takes $I4/mmm$ symmetry when the pressure exceeds 19 GPa. 
	La$_4$Ni$_3$O$_{10}$ has $P2_1/a$ symmetry under normal conditions. However, applying pressure greater than 12.6 to 13.4 GPa, even at ambient temperatures, induces a transition to $I4/mmm$ symmetry.
	Both La$_3$Ni$_2$O$_7$ and La$_4$Ni$_3$O$_{10}$ have  $I4/mmm$ symmetry in the superconducting phase. La$_4$Ni$_3$O$_{10}$ features three Ni-O layers, whereas La$_3$Ni$_2$O$_7$ has two. 
	$I4/mmm$ La$_4$Ni$_3$O$_{10}$ contains two nonequivalent nickel atoms, Ni1 and Ni2, as illustrated in Fig.\ref{fig:4310_327_structure}(a). 
	In contrast, La$_3$Ni$_2$O$_7$ consists of only one type of nickel atom, as shown in Fig.\ref{fig:4310_327_structure}(b). 
	Ni1 and Ni2 have distinct properties, with interlayer doping.	
	It is to be noted that the La-O layers between the NiO layers have net positive charge based on the valence.
	
	We study electronic properties of high-pressure structures from 15.3 GPa to 44.3 GPa reported in Ref.\cite{10.1007/s11433-023-2329-x}. 
	We used the efficient APW+lo method \cite{sjostedt} as implemented in the WIEN2k package\cite{10.1063/1.5143061}, 
	and the Perdew-Burke-Ernzerhof (PBE) generalized gradient approximation (GGA).
	We used $30\times30\times30$ {\bf k}-point meshes
were used to sample the Brillouin zone during the self-consistent iterations. 
	The muffin-tin radii in Bohr were  Ni:1.82,La:2.2,O:1.56 and
the basis set cutoff parameter was $R_{min}K_{max}$=7. 
	Both muffin-tin radii and $R_{min}K_{max}$ were fixed in all calculations in order to facilitate comparisons.
	
\begin{figure*}[htbp]
		\centering
		\includegraphics[width=\textwidth]{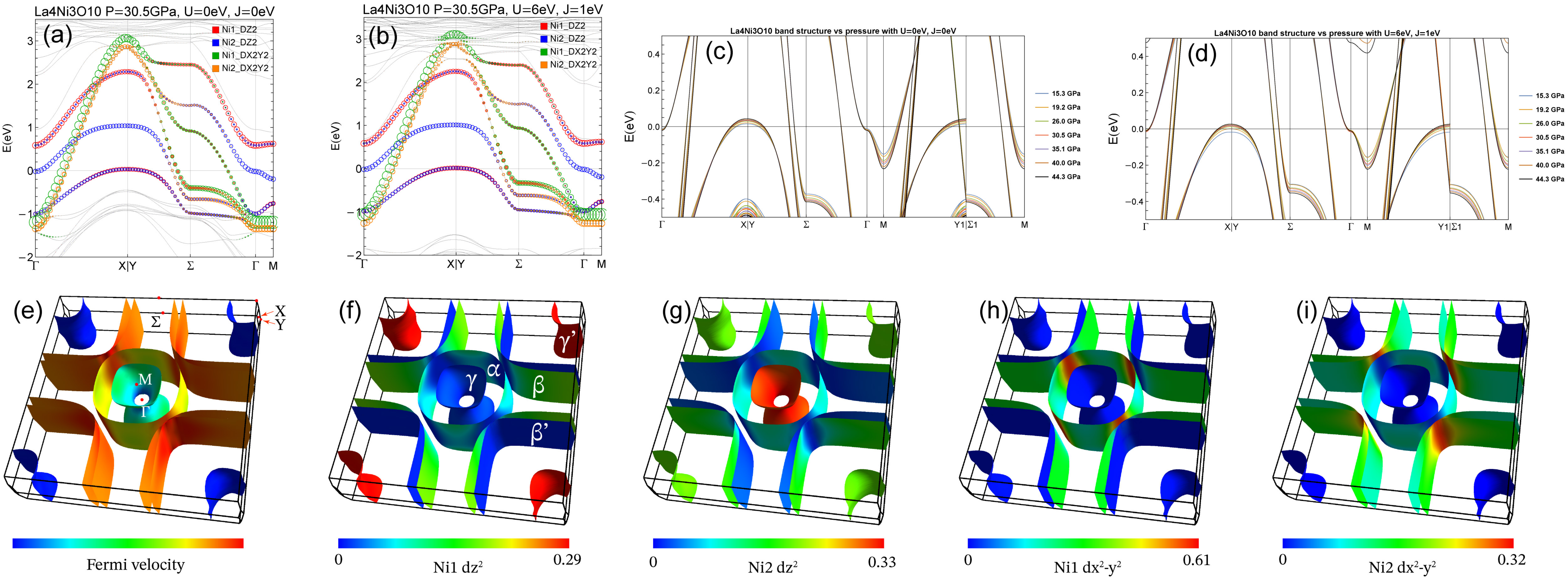}
		\caption{(a)(b)(f)(g)(h)(i) are orbital resolved band and Fermi surface of La$_4$Ni$_3$O$_{10}$ at
30.5 GPa without $U$ and $J$ for Ni1 $d_{z^2}$, Ni2 $d_{z^2}$, Ni1 $d_{x^2-y^2}$ and Ni2 $d_{x^2-y^2}$ respectively. Note that the color scale for each Fermi surface is different. Also, the orbital weight does not count atom multiplicity.
			In (f), we denote 5 Fermi surface sheets as $\alpha$,$\beta$,$\beta'$,$\gamma$,$\gamma'$
			(e) Fermi velocity with Brillouin zone high symmetry point notation used in this paper,
			(c) bands under different pressure plotted together. Note that
the lengths of plotted k-path segments under different pressures are all normalized to be the same as those at 30.5 GPa
although the lattice parameters are different,
			(d) same as (c) but with $U$=6 eV and $J$=1 eV
		}
		\label{fig:big_compose}
	\end{figure*}

	The octahedral crystal field leads to $e_g$ orbitals near the Fermi level in these compounds, specifically the $d_{x^2-y^2}$ and $d_{z^2}$ orbitals, which are hybridized with in-plane and apical O $p$ orbitals, respectively. 
	In Fig.\ref{fig:big_compose}, we show La$_4$Ni$_3$O$_{10}$ at 30.5 GPa as an example.
We show orbital resolved bands and Fermi surfaces for Ni1 $d_{z^2}$, Ni2 $d_{z^2}$, Ni1 $d_{x^2-y^2}$ and Ni2 $d_{x^2-y^2}$. 
	The high symmetry points of BZ are labeled in Fig.\ref{fig:big_compose}(e). The Fermi surfaces are labeled in Fig.\ref{fig:big_compose}(f) as $\gamma,\gamma',\alpha,\beta,\beta'$
	
	In Fig.\ref{fig:big_compose}(b), we applied nonzero Hubbard $U$=6 eV and Hund $J$=1 eV compared to Fig.\ref{fig:big_compose}(a). 
	Surprisingly, the $d_{z^2}$ and $d_{x^2-y^2}$ bands are little changed in terms of the band width or orbital occupancy with applied $U$ and $J$. The main effect of $U$ and $J$ is pushing $t_{2g}$ bands
significantly downwards to higher binding energy. Here we emphasize results that do not depend on the choice of $U$.
	In Fig.\ref{fig:big_compose}(c) and (d), we plot bands under different pressure together and focus on the range
$E_{F}$-0.5 eV to $E_{F}$+0.5 eV. 
	We can see that pressure has a significant impact on the $\gamma'$ pocket; as the pressure increases, the $\gamma'$ pocket becomes larger. In addition, $U$ and $J$ mainly affect the $\gamma'$ pocket, especially with relatively low pressures of 15.3 GPa and 19.2 GPa, where applying $U$ and $J$ suppresses the $\gamma'$ pocket. Under higher pressures, $U$ and $J$ also shrink the $\gamma'$ pocket. The other Fermi pockets, however, are insensitive to both pressure and $U$, $J$. 
	Since the influences of $U$, $J$, and pressure on the energy bands near the Fermi level are minimal, our following discussion is based on the bands where $U$ and $J$ are zero, and the pressure is 30.5 GPa.
	
	From Fig.\ref{fig:big_compose}(a), we can see clear separation of $d_{z^2}$ and $d_{x^2-y^2}$ on $\Gamma-M$ and $\Gamma-X$.
	$\Gamma-M$ is in $kz$ direction, and  $d_{x^2-y^2}$ is flat along this direction, but $d_{z^2}$ has dispersion.
	The $\gamma$ pocket at $\Gamma$ point is from Ni2 $d_{z^2}$ without any Ni1 component as shown in Fig.\ref{fig:big_compose}(a)(g) on the whole $\Gamma-M$ and $\Gamma-X$ path. 
	As seen from Fig.\ref{fig:Ni_vs_O_band}(d), the $p_z$ orbitals
of both the O2 and O4 contribute to this band. 
	A similar situation is found on $\gamma'$ pocket at X point. 
	The $\gamma'$ pocket contributed by $d_{z^2}$ of both Ni1 and Ni2, with Ni1 contributing slightly more. However, considering that there are actually two Ni2 atoms, overall, Ni2 contributes more. 
	However, as in Fig.\ref{fig:Ni_vs_O_band}(d), there is no contribution from O2's $p_z$ orbital at X point.
	The $\beta'$ pocket is contributed by Ni2's $d_{x^2-y^2}$ and $d_{z^2}$, mainly $d_{x^2-y^2}$. However, $\beta$ pocket is a mix of Ni2's $d_{x^2-y^2}$ and Ni1's $d_{z^2}$ and $d_{x^2-y^2}$
	The $\alpha$ pocket is contributed by all four orbitals, but most weight concentrated on $\Gamma-X$ direction by Ni1's $d_{x^2-y^2}$

	\begin{figure*}[htbp]
		\centering
		\includegraphics[width=\textwidth]{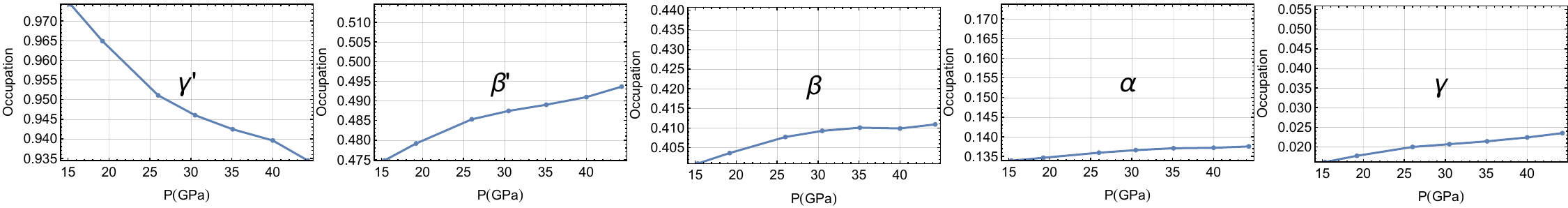}
		\caption{Band occupancy vs pressure for the five Fermi surface sheets of La$_4$Ni$_3$O$_{10}$ without $U$ and $J$. The vertical range of all plots are set the same in order for comparison. }
		\label{fig:occupation}
	\end{figure*}
	We present plots of electron occupancy vs pressure for different Fermi surfaces of La$_4$Ni$_3$O$_{10}$ in Fig.\ref{fig:occupation}. As pressure increases, the $\gamma'$ band exhibits a noticeable decrease in occupancy, indicating electron depletion, due to electrons being transferred to other bands. Conversely, the $\beta'$ and $\beta$ bands show substantial increases in occupancy, suggesting that these bands are being significantly electron-doped under pressure. 
	This indicates a pronounced transfer of electrons into these bands from the $\gamma'$ band. 
	The $\alpha$ and $\gamma$ bands exhibit slight increases in occupancy with increasing pressure, although these changes are less pronounced compared to the $\beta'$ and $\beta$ bands. 
	This suggests a minor self-doping effect in the $\alpha$ and $\gamma$ bands. Overall, the pressure-induced self-doping effect in La$_4$Ni$_3$O$_{10}$ results in a redistribution of electrons among the Fermi surface sheets, with the $\beta'$ and $\beta$ bands having increased electron occupancy coming from the $\gamma'$ band.
	
	We emphasize the similarity between infinite layer NdNiO$_2$ doped with strontium and cuprates, for example, their Fermi surfaces are very similar\cite{10.1103/PhysRevLett.125.077003}. 
	Magnetic modes observed by resonant inelastic x-ray scattering are also similar to those of the typical doped Mott insulator in cuprates\cite{10.1126/science.abd7726}. 
	However, in La$_3$Ni$_2$O$_7$ Ni has valence Ni$^{2.5+}$, quite different from cuprates. 
	Nonetheless, recent experiments\cite{10.1038/s41567-024-02515-y} show superconductivity similar to cuprates and features such as linear temperature-dependent resistivity.
We emphasize
with the different Ni the electron count is very different from cuprates but
our calculations show the $\beta$ and $\beta'$ Fermi surfaces in La$_4$Ni$_3$O$_{10}$ remain similar shape to those in cuprates. 
	However, the differences are also significant. 
	Both cuprates and doped NdNiO$_2$ have main Fermi surface sheets primarily from the $d_{x^2-y^2}$ orbitals hybridized with in-plane O$_p$ orbitals.
	The $\beta$ and $\beta'$ Fermi surfaces, also dominantly from $d_{x^2-y^2}$ orbitals,
with hybridization by both in-plane O, and also the $d_{z^2}$ and apical O are analogous
both in shape and orbital composition to the superconducting sheets in the cuprates and doped NdNiO$_2$.
The $\beta'$ sheet has occupancy close to half-filling while $\beta$ sheet is doped away from half-filling by approximately $0.09\sim0.10$ holes,
depending on pressure. Both $\beta$ and $\beta'$ have Ni2 $d_{x^2-y^2}$ character,
and in the case of $\beta$ there is additionally Ni1 $d_{x^2-y^2}$ character almost twice that of
Ni2 $d_{x^2-y^2}$ per atom. The strongest effect of pressure is an increase in electron count towards
half-filling with pressure for the $\beta$ sheet.

		\begin{figure*}[htbp]
		\centering
		\includegraphics[width=\textwidth]{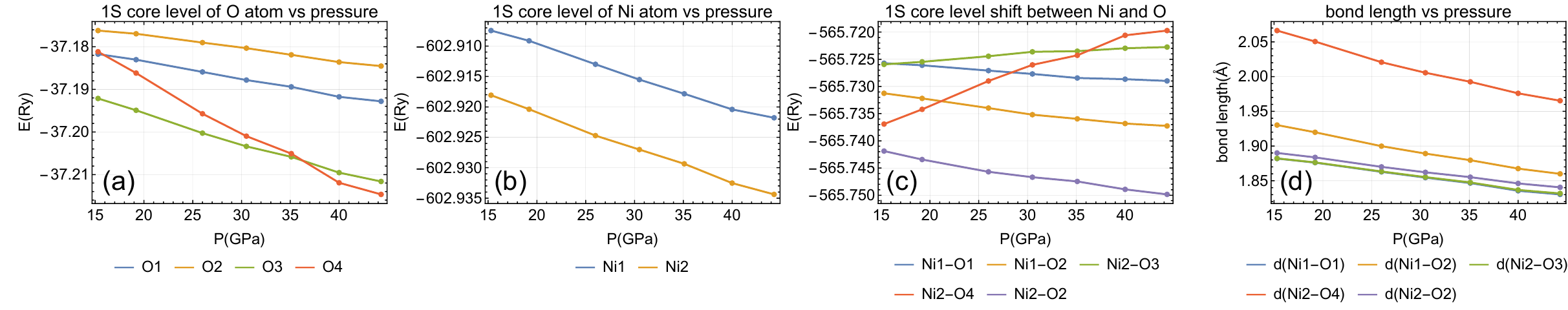}
		\caption{
			(a) 1$s$ core level of O1, O2, O3, O4 relative to $E_F$ vs pressure,
			(b) 1$s$ core level of Ni1, N2 relative to $E_F$ vs pressure.
			(c) differences between various Ni and O 1$s$ core levels vs pressure.
			(d) bond length of various Ni-O vs pressure.
		}
		\label{fig:core_level_and_bond_info}
	\end{figure*}

	We now turn to the connections between structure and the doping of the sheets.
	In Fig.\ref{fig:core_level_and_bond_info}(d), we plot variations of various Ni-O bonds vary with pressure. 
	In Fig.\ref{fig:core_level_and_bond_info}(a)(b)(c), we plot 1$s$ core level relative to $E_F$ for oxygen and nickel sites and their differences vs pressure. 
	As expected, all Ni-O bond lengths decrease with pressure. The planar bond Ni1-O1 and Ni2-O3 are the shortest. The next longer bonds are Ni2-O2 and Ni1-O2 in which O2 bridges Ni1 and Ni2. 
	Finally, the longest bond is Ni2-O4, which is the outer apical O bond of the nickel trilayer.
	Ni2 octahedra has Ni2-O2 and Ni2-O4 with different lengths. 
	The 1$s$ core level of Ni2 is lower than Ni1 indicating their different environment.
	Both the Ni1 and Ni2 core levels decreases with increasing pressure.
	Interestingly, despite Ni2-O4 being the longest bond, under low pressure, both the 1$s$ core level of O4 and the core level difference between Ni2 and O4 are at intermediate values. 
	As the pressure increases to the maximum, the 1$s$ core level of O4 decreases faster than those of other O atoms and reaches the lowest value, while the core level difference between Ni2 and O4 goes up above all other core level differences under pressure.

	Thus, the Ni2-O4 bond, which represents the behavior of the outer layer apical O is distinct. The downward shift indicates a stabilization of the ionic nature of this outer apical O with pressure. It is to be noted that the crystal field split $e_g$ orbitals are antibonding states. Therefore, the consequence of the shortening of the softer Ni2-O4 bond relative to the others is expected to push bands associated with the Ni2 $d_{z^2}$ orbital to higher energy. Hence, the reduced occupation under pressure of the $\gamma'$ Fermi surface.
	
	Therefore, under pressure electrons are transferred from the outer-layer Ni2 $d_{z^2}$ of $\gamma'$ Fermi surface to the $\beta$ and $\beta'$ sheets, which have mixed Ni1 and Ni2 $d_{x^2-y^2}$ character. This represents an unconventional interlayer doping, with transfer from a Fermi surface originating on the outer layer, to Fermi surfaces from both the inner and outer layer. The pressure dependent occupations are particularly interesting in relation to a scenario where the $\beta$ and $\beta'$, $d_{x^2-y^2}$ sheets are the active sheets for superconductivity.

Firstly, it may be noted that there are only two such sheets, while there are three Ni-O layers. This is a consequence of the band formation, which pushes one of the $d_{x^2-y^2}$ sheets to higher energy as shown in Fig. \ref{fig:big_compose}.
This, plus the presence of other sheets of Fermi surface, representing partially occupied bands,
accounts for the valence difference from the Ni$^{1+}$ of NdNiO$_2$,
while still maintaining the cuprate-like structure of the $\beta$ and $\beta'$ surfaces.
Secondly, the $\beta'$ sheet is near half filling, and becomes increasingly close to half filling with pressure.
In cuprates, half-filling is associated with a Mott insulating state, incompatible with superconductivity.
Doping away from half-filling has at least two effects.
The first is destruction of the Mott insulating state in favor of a conducting state, with Fermi surfaces,
compatible with superconductivity.

The second is suppression of antiferromagnetism with reduction of the corresponding spin-fluctuations as the hole doping is increased, particularly towards the over-doped region of the phase diagram. In present case, due to the multi-orbital nature of the bands and the multiple bands crossing $E_F$ there is no half-filled orbital that would favor a Mott insulating state consistent with the fact that La$_4$Ni$_3$O$_{10}$ is not insulating. However, the cuprate-like near half-filled $\beta'$ may lead to nearness to antiferromagnetism, with the doping level of this band being a key parameter. This suggests studies probing the proximity to magnetism of this compound both in terms of the O stoichiometry and pressure, especially in relation to the superconducting properties.
	
	Thus, in spite of the very different electron counts the superconductivity
of the different nickelate superconductors can be unified in terms of the Fermi surface structure,
which universally shows similar $d_{x^2-y^2}$ sheets characteristic also of the cuprate high-Tc materials.
This provides a framework that enables similar superconductivity in cuprates and nickelates.
In addition, by avoiding the Ni$^{1+}$ valence state and the chemical instability of this state,
but maintaining an electronic structure favorable for superconductivity materials like La$_3$Ni$_2$O$_7$ and La$_4$Ni$_3$O$_{10}$ point to the possible existence of many more layered superconducting nickelates that remain to be discovered.

	Y. N. Huang thanks W. C. Bao and H. Q. Lin for helpful discussions. 
	Y. N. Huang is supported by the National Natural Science Foundation of China (Grant No. 11904319).

	\bibliography{La4Ni3O10}

%%%%%%%%%% Merge with supplemental materials %%%%%%%%%%
\pagebreak
\widetext
\begin{center}
	\textbf{\large Supplemental Materials: Unconventional Pressure Dependent Interorbital and Interlayer Doping in Superconducting Nickelates}
\end{center}
%%%%%%%%%% Merge with supplemental materials %%%%%%%%%%
%%%%%%%%%% Prefix a "S" to all equations, figures, tables and reset the counter %%%%%%%%%%
\setcounter{equation}{0}
\setcounter{figure}{0}
\setcounter{table}{0}
\setcounter{page}{1}
\makeatletter
\renewcommand{\theequation}{S\arabic{equation}}
\renewcommand{\thefigure}{S\arabic{figure}}
\renewcommand{\bibnumfmt}[1]{[S#1]}
\renewcommand{\citenumfont}[1]{S#1}
%%%%%%%%%% Prefix a "S" to all equations, figures, tables and reset the counter %%%%%%%%%%

	\begin{figure*}[htbp]
	\centering
	\includegraphics[width=0.8\textwidth]{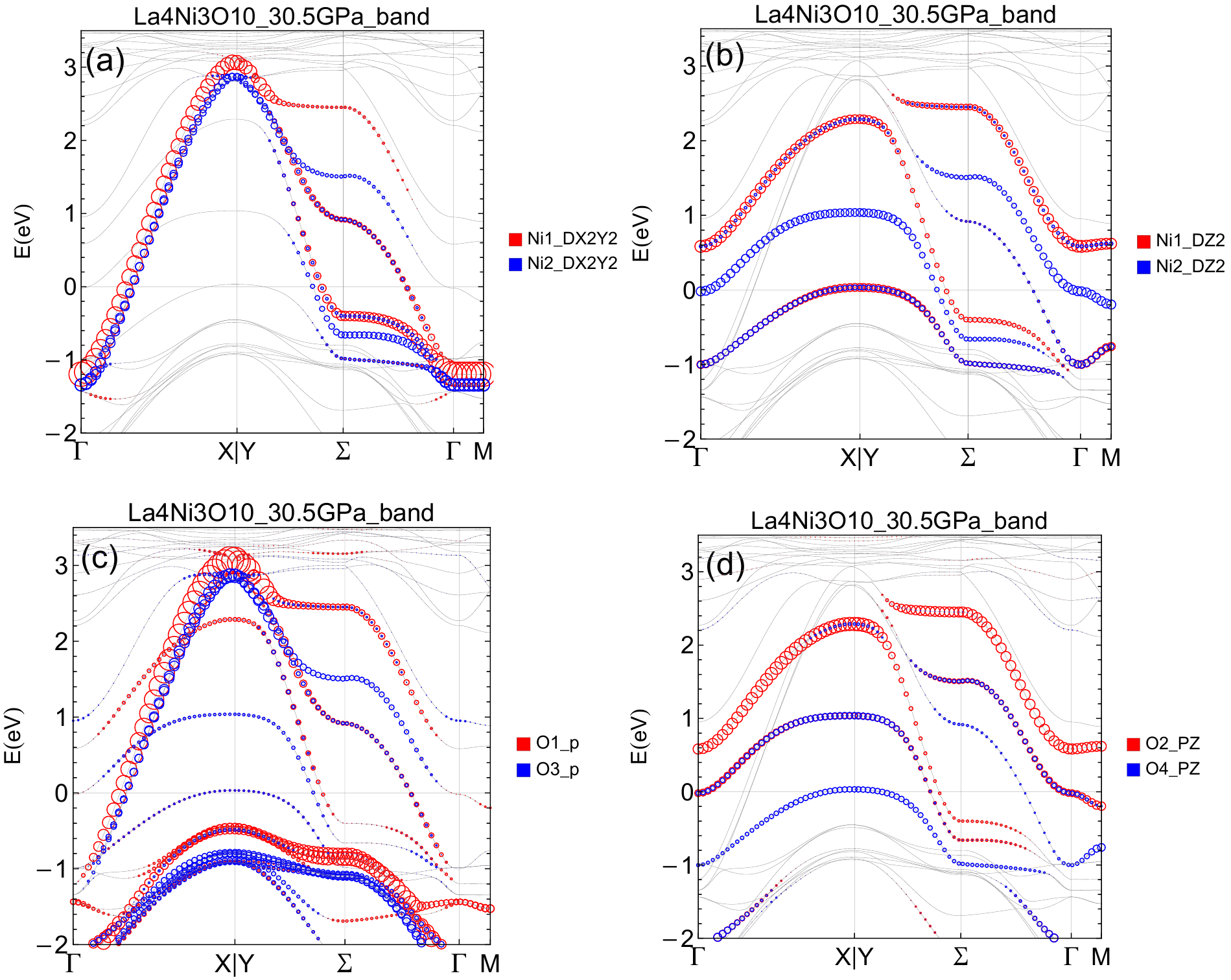}
	\caption{
		(a) projected $d_{x^2-y^2}$ bands of Ni1 and Ni2 of La$_4$Ni$_3$O$_{10}$,
		(b) projected $d_{z^2}$ bands of Ni1 and Ni2 of La$_4$Ni$_3$O$_{10}$,
		(c) projected $O_p$ bands of equatorial O1 and O3 of La$_4$Ni$_3$O$_{10}$, 
		(d) projected $O_{pz}$ of apical O2 and O4 bands of La$_4$Ni$_3$O$_{10}$. 
		Note that the scale of O projected bands are multiplied by 5 compared to Ni projected bands.
	}
	\label{fig:Ni_vs_O_band}
\end{figure*}

\end{document}